\documentclass[%
reprint,
superscriptaddress,
preprintnumbers,
nobibnotes,
nofootinbib,
 amsmath,amssymb, 
 aps, prd,
 longbibliography,
]{revtex4-1}

\usepackage{cancel}
\usepackage{accents}
\usepackage{mciteplus,slashed}
\usepackage{amssymb,cancel,amsmath}
\usepackage{dcolumn}
\usepackage{bm}
\usepackage{soul}
\usepackage[caption=false]{subfig}
\usepackage{appendix}
\usepackage{physics}
\usepackage{booktabs}
\usepackage{comment}
\usepackage[T1]{fontenc}	
\usepackage{csvsimple}
\usepackage[bookmarksnumbered=true]{hyperref} 
\hypersetup{
    colorlinks = true,
    linkcolor = blue,
    anchorcolor = blue,
    citecolor = blue,
    filecolor = blue,
    urlcolor = blue
    }
\usepackage[section]{placeins}
\usepackage[capitalise]{cleveref}
\usepackage{booktabs}
\usepackage{graphicx}
\usepackage{mathrsfs}
\usepackage{relsize}
\graphicspath{{Figures/}}
\AtBeginDocument{
\heavyrulewidth=.08em
\lightrulewidth=.05em
\cmidrulewidth=.03em
\belowrulesep=.65ex
\belowbottomsep=0pt
\aboverulesep=.4ex
\abovetopsep=0pt
\cmidrulesep=\doublerulesep
\cmidrulekern=.5em
\defaultaddspace=.5em
}
\usepackage[dvipsnames]{xcolor}
\usepackage[normalem]{ulem}
\usepackage{fontawesome} 
\usepackage[force]{feynmp-auto}
\usepackage{bbm}
\usepackage{MnSymbol}
\usepackage{comment}
\usepackage[force]{feynmp-auto}

\def\iu{\mathrm{i}}
\def\e{\mathrm{e}}

\def\hyd{~\!\!^1{\rm H}}

\definecolor{forestgreen}{HTML}{228B22}

\begin{document}

\title{Final state interactions for high energy scattering off atomic electrons}

\author{Ryan Plestid}
\author{Mark B.\ Wise}
\affiliation{Walter Burke Institute for Theoretical Physics, California Institute of Technology, Pasadena, CA 91125, USA}

\date{\today}

\preprint{CALT-TH/2024-026}

\begin{abstract}
    We consider the scattering of high energy leptons off bound atomic electrons focusing primarily on final state interactions 
    between the outgoing energetic electron, and the heavy residual charged ``debris'' in the final state. These effects are inherently absent from calculations for a free electron at rest. Coulomb exchanges are enhanced by the large number of electrons in the atomic debris, and are unsuppressed by non-relativistic velocities in the debris. We find that these exchanges can be resummed using operator methods, and cancel at the level of the cross section until at least $O(\alpha^3)$. Furthermore, we argue that both final {\it and} initial state Coulomb exchanges (enhanced by the number of electrons in the atom) do not affect the cross section until at least $O(\alpha^3)$. Transverse photon couplings  to non relativistic electrons are proportional to their small velocities, and rotational invariance suppresses their contribution to $O(\alpha^3)$.   Our results are relevant for precision experiments involving  neutrinos, electrons, positrons, and muons scattering off of atomic electrons in a fixed target. 
%

\end{abstract}

 \maketitle 

\section{Introduction \label{Intro} }

Fixed target experiments play an important role in high energy physics at the intensity frontier. Examples include neutrino experiments \cite{Nunokawa:2007qh,Hyper-KamiokandeWorkingGroup:2014czz,Diwan:2016gmz,Acero:2019ksn,T2K:2021xwb,DUNE:2020ypp}, searches for feebly interacting particles \cite{Berlin:2018bsc,Nardi:2018cxi,Marsicano:2018glj,Arias-Aragon:2024qji,NA64:2023wbi}, rare processes \cite{Perrevoort:2018ttp,Bernstein:2019fyh,KOTO:2020prk,MEGII:2021fah,NA62:2021zjw,REDTOP:2022slw}, and precision measurements \cite{SLACE158:2005uay,MOLLER:2014iki,Abbiendi:2016xup,PiENu:2015seu,PIONEER:2022yag}. It is often implicitly assumed that atomic electrons can be treated as free and at rest when they participate in a scattering interaction with an incident high energy beam. This is a good leading approximation, however for high precision experiments sub-leading corrections must be understood theoretically. 

In recent work \cite{Plestid:2024xzh} we have studied the corrections due to atomic binding arising both from phase space, and the initial state (many-body) wavefunction of the atom. These corrections stem from the finite three-momentum of electrons in bound atomic orbitals, and the shifted kinematics from binding energies. Both of these effects enter at the same order and they must be treated simultaneously. They account for the non-perturbative bound-state dynamics i.e., from iterated Coulomb exchange in the initial state. 

Further corrections arise from perturbative photon exchange between ``hard'' leptons, and ``soft'' (i.e., non-relativistic) electrons and nuclei. These effects are inherently absent from calculations for a free electron at rest. For example, for neutrino scattering off an atomic electron, the outgoing $\sim {\rm GeV}$ electron can exchange photons with the ionized ``atomic debris''. Incident charged lepton scattering (e.g., $\mu e \rightarrow \mu e$) involves both initial and final state photon exchange with atomic electrons or nuclei.

In this work, we focus mostly on perturbative final-state interactions involving the outgoing energetic electron (in \cref{Applications} we also comment on the ``shake-up'' of the initial state from Coulomb exchange with an incoming charged lepton). We specifically focus on ladder-graphs involving photon exchanges with the final state debris. If we track the charge of the debris, $Q_B$, and the charge of the electron, $Q_e$, separately then the (gauge invariant) class of corrections we consider are of order\footnote{We explicitly neglect ``mixed'' corrections such as those of order $O(Q_BQ_e^3)$.} $(Q_B Q_e)^n$.  Although $Q_B=+e$, matrix elements involving the debris are enhanced for multi-electron atoms. This enhancement occurs for momentum transfers on the order of the atomic scale (i.e., the inverse Bohr radius) when the struck electron is able to see a partially (but not fully) screened atomic nucleus.  To match the order studied in Ref.~\cite{Plestid:2024xzh}, requires working to order $(Q_B Q_e)^2$.

These corrections are a necessary input for an extraction of hadronic vacuum polarization from $\mu e \rightarrow \mu e$ scattering \cite{CarloniCalame:2015obs} as is being pursued by the MuonE collaboration \cite{Abbiendi:2016xup,Abbiendi:2022oks}. For neutrino electron scattering (relevant for neutrino flux normalization measurements \cite{Marshall:2019vdy}), we find these corrections are much smaller than the uncertainty from the 1-loop radiative corrections computed in Ref.~\cite{Tomalak:2019ibg} (roughly $\sim 0.5\%$).

We use a non-relativistic power counting for photon interactions with the atomic system, and standard perturbative counting for the highly relativistic electron in the final state. We find a number of interesting results. First, Coulomb corrections cancel in the differential cross section upon integration over the final state phase space of the atomic debris; this result holds to all orders in perturbation theory [i.e., resumming graphs of order $(Q_B Q_e)^n$] in the limit of small energy splittings of the debris. For hydrogen, the cancellation can be understood in terms of a final-state wavefunction, whereas for multi-electron atoms it stems from an operator-level identity. Second, sub-leading corrections due to a single transverse photon exchange in are also shown to cancel. Finally, a class of corrections which are sensitive to the energy splittings of the atomic debris and appear at order $e^2$ can be shown to have vanishing interference with the leading order matrix element. Taken together these results imply that final state interactions (as defined above) do not alter the predictions of Ref.~\cite{Plestid:2024xzh}  up to and including $O(Q_e^2Q_B^2)$ in our power counting. A similar result holds for interactions that involve the initial atomic state.

In everything that follows, we will omit factors of $Q_e$ and $Q_B$ and simply count powers of $\alpha$. This includes factors of $\alpha$ that stem from bound-state dynamics i.e., that the typical velocity of an electron in an atom is $O(\alpha)$. We aim to work to an accuracy of $O(\alpha^2)$. 


The rest of the paper is organized as follows: In \cref{Hydrogen} we discuss the case of scattering neutrinos from hydrogen. We find that Coulomb distortion of the final state cancels at the level of the cross section after integrating over the final-state proton's phase space. Next in \cref{Multi_Electron} we show how to generalize our analysis to multi-electron atoms where the corrections similarly cancel at the level of the cross section. We find an operator level expression for Coulomb photons, and also discuss sub-leading corrections such as transverse photon exchange with non-relativistic electrons in the final state debris. Transverse photons are discussed in \cref{Trans_Photon}. We find the non-relativistic nature of the electrons combined with rotational invariance of the initial atom implies that transverse photons first enter at $O(\alpha^3)$. In \cref{Applications} we describe how to apply our results at the level of a measurable cross section. We show that the relevant sum-rules that control cross sections (inclusive with respect to the debris)  are unaffected by Coulomb exchange through $O(\alpha^2)$. We discuss how this result applies both to neutrino electron scattering and muon electron scattering.  
Finally in \cref{Conclusions} we provide a summary and outlook.

\section{Hydrogen \label{Hydrogen} }
Let us first focus on neutrino scattering off hydrogen as the simplest example where only final state interaction are present, 
\begin{equation}
    \nu(\vb{k}) + \hyd(\vb*{0}) \rightarrow \nu(\vb{k}') + e^-(\vb{p}') + p^+(\vb{h}')~. 
\end{equation}
The outgoing proton is heavy, and couples dominantly to Coulomb modes with couplings to transverse photons suppressed by $|\vb{h}'|/m_p\sim \alpha m_e/m_p$. While discussing hydrogen, we will therefore focus exclusively on the exchange of Coulomb photons in the final state. 

Consider an outgoing scattering state $\ket{f}=\ket{e^-(\vb{p}') p^+(\vb{h}')}$ labeled by the asymptotic momentum of the electron and proton. Including the iterated Coulomb potential, the state $\ket{f}$ satisfies a Lippmann-Schwinger equation 
\begin{equation}
    \ket{f} = \ket{e^-(\vb{p}')}_0 \ket{p^+(\vb{h}')}_0 + \frac{1}{E_f - \hat{H}_0 -\iu 0} \hat{V} \ket{f}~, 
\end{equation}
where $\ket{e^-(\vb{p}')}_0\ket{p^+(\vb{h}')}_0$ is the free-particle state. 
Here $\hat{V}$ is the Coulomb interactions, $\hat{H}_0$ is the free-Hamiltonian and $E_f = \sqrt{\vb{p}'^2+m_e^2} + \sqrt{\vb{h}'^2+m_p^2}$ is the exact energy of $\ket{f}$; the causal regulator ($-\iu 0$) is fixed by demanding that $\ket{f}$ be an out-state. 
The final state can be written in terms of its valence free-particle Fock states, as\footnote{Strictly speaking the Dirac wavefunction should be spin-dependent, but we suppress spin indices for simplicity. The answer is unaffected in the kinematic regime of interest \cite{Tjon:2006qe} and \cref{final-state} is correct as written.}
\begin{widetext}
\begin{equation}
    \begin{split}
    \label{final-state}
    \ket{f(\vb{p}',\vb{h}')} = \sqrt{2E_e({\vb{p}'})} \sqrt{2 E_p({\vb{h}'})} &\int \frac{\dd^3 q}{(2\pi)^3} \frac{\tilde{\phi}_{\vb{p}',\vb{h}'}(\vb{q})}{\sqrt{2E_e({\vb{p}'+\vb{q}})}\sqrt{2E_p({\vb{h}'-\vb{q}})}} \ket{e^-(\vb{p}'+\vb{q})}_0\ket{p^+(\vb{h}'-\vb{q})}_0~. 
    \end{split}
\end{equation}
\end{widetext}
Note that when $\hat{V}=0$, the scattering wavefunction is $\tilde{\phi}_{\vb{p}',\vb{h}'}(\vb{q})= (2\pi)^3 \delta^{(3)}(\vb{q})$. The two-particle scattering state is normalized according to 
\begin{equation}
    \begin{split}
    \label{state-norm} 
    \braket{f(\vb{p}',\vb{h}')}{f(\vb{p},\vb{h})} =  2&E_e(\vb{p}) (2\pi)^3 \delta^{(3)}(\vb{p}-\vb{p}') \\
    & \times 2 E_p(\vb{h})(2\pi)^3 \delta^{(3)}(\vb{h}-\vb{h}')~. 
        \end{split}
\end{equation}
The scattering state's wavefunction depends only on the asymptotic relative momentum $\vb{f}'\equiv \vb{p}'-\vb{h}'$ so that $\tilde{\phi}_{\vb{p}',\vb{h}'}(\vb{q}) = \tilde{\phi}_{\vb{f}'}(\vb{q})$.  The scattering state wavefunction is normalized according to 
\begin{equation}
    \begin{split}
     \label{wf-norm}
    \int& \frac{\dd^3 q}{(2\pi)^3} \tilde{\phi}_{\vb{f}'}^*(\vb{q}+\vb{f}') \tilde{\phi}_{\vb{f}}(\vb{q}+\vb{f})= (2\pi)^3 \delta^{(3)}(\vb{f}'-\vb{f})~.
    \end{split}
\end{equation}
It is readily checked that \cref{wf-norm} can be substituted into \cref{final-state} to obtain \cref{state-norm}. In coordinate space \cref{wf-norm} becomes, 
\begin{equation}
    \begin{split}
     \label{wf-norm-x}
    \int& \dd^3 x ~ \e^{\iu (\vb{f}-\vb{f}')\cdot \vb{x}}~\phi_{\vb{f}'}^*(\vb{x}) \phi_{\vb{f}}(\vb{x})= (2\pi)^3 \delta^{(3)}(\vb{f}'-\vb{f})~.
    \end{split}
\end{equation}

We similarly write the initial hydrogen atom in terms of free-particle valence states, 
\begin{equation}
    \begin{split}
    \!\ket{\hyd(\vb*{0})}= \!\int\!\!\!\frac{\dd^3 p}{(2\pi)^3}
    &\frac{\sqrt{2M_H}}{\sqrt{2E_e(\vb{p})}\sqrt{2 E_p(\vb{p})}} \\
    &\hspace{0.1\linewidth}
   \times\tilde{\psi}(\vb{p})\ket{e^-(\vb{p})}_0 \ket{p^+(-\vb{p})}_0~.
    \end{split}
\end{equation}
The bound state's wavefunction is normalized according to 
\begin{equation}
    \int \frac{\dd^3q}{(2\pi)^3} |\tilde{\psi}(\vb{q})|^2=1~.
\end{equation}
The relevant matrix element is 
\begin{equation}
    \mathcal{M} = \mel{f(\vb{p}',\vb{h}') \nu(\vb{k}')}{\mathcal{H}_W}{\hyd(\vb{0})\nu(\vb{k})}~,
\end{equation}
where $\mathcal{H}_W$ is the weak Hamiltonian density evaluated at the origin of spacetime. Using the above definitions, and $\!_0\braket{p^+(\vb{h}'-\vb{q})}{p^+(\vb{-p})}_0= 2 E_p (2\pi)^3 \delta^{(3)}(\vb{q}-(\vb{h}'+\vb{p}))$, we arrive at 
\begin{equation}
    \label{matrix-element-expression}
    \begin{split}
    \mathcal{M} &= \int \frac{\dd^3 p}{(2\pi)^3} \sqrt{\frac{2M_H 2 E_p}{2E_e}} \tilde{\phi}_{\vb{p}'-\vb{h}'}^*(\vb{p}+\vb{h}') \tilde{\psi}(\vb{p}) \\
    &\hspace{0.35\linewidth}\times {\sf M}(\vb{p}'+\vb{p}+\vb{h}',\vb{k}';\vb{p},\vb{k}) \\
    &\simeq \int \frac{\dd^3 p}{(2\pi)^3} \sqrt{\frac{2M_H 2 E_p}{2E_e}}  \tilde{\phi}_{\vb{p}'-\vb{h}'}^*(\vb{p}+\vb{h}') \tilde{\psi}(\vb{p}) \\
    &\hspace{0.45\linewidth}\times {\sf M}(\vb{p}',\vb{k}';\vb{p},\vb{k})~,
    \end{split}
\end{equation}
where in the second line we have neglected the $\vb{p}+\vb{h}'$ (both small) relative to $\vb{p}'$ (large); the error due to this approximation is very small $\sim |\vb{p}|/|\vb{p}'|$ \cite{Plestid:2024xzh}. The free-electron matrix element is defined as
\begin{equation}
    {\sf M}(\vb{p}',\vb{k}';\vb{p},\vb{k})= \!~_0\langle e^-(\vb{p}')\nu(\vb{k}')| \mathcal{H}_W| e^-(\vb{p})\nu(\vb{k})\rangle_0~. 
\end{equation}
As mentioned above, in the absence of Coulomb exchanges $\tilde{\phi}_{\vb{p}'-\vb{h}'}(\vb{q}) \rightarrow (2\pi)^3 \delta^{(3)}(\vb{q})$, and \cref{matrix-element-expression} reduces to Eq.~(14) of Ref.~\cite{Plestid:2024xzh}.
\subsection{Perturbation theory \label{Pert_Theory} }
\begin{figure}
    \centering
    \includegraphics[width=0.9\linewidth]{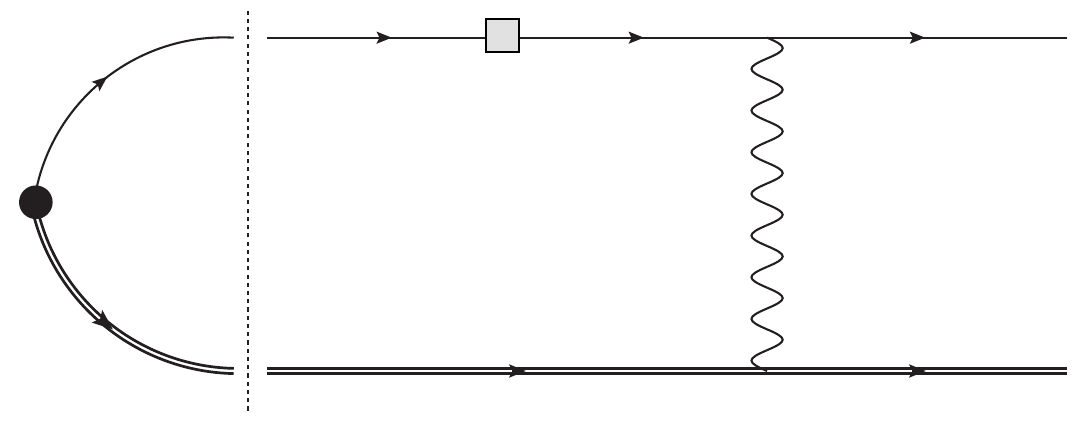}
    \caption{Diagramatic representation of the $O(\alpha)$ correction to the final state $\ket{f}$ from the exchange of a Coulomb mode. One computes the tree-level diagram to the right of the dashed line and folds it against the wavefunction to the left of the dashed line. The proton is drawn as a double line, the electron as a single line, the neutrino-electron vertex as a grey box, and the photon as a wavy line.}
    \label{fig:marks_diagram}
\end{figure}
It is instructive to compute the matrix element perturbatively. This both serves to illustrate the size of Coulomb corrections at the level of the amplitude and the connection between diagramatics and wavefunctions. Furthermore, by analyzing different regions of loop momenta we learn which characteristic scales dominate the Coulomb corrections. 

The $O(\alpha)$ correction involves an integral of the form (including a photon mass, $\mu_\gamma$, to regulate IR divergences)
\begin{equation}\label{1-loop}
   I^{(1)}=\! \int\!\!  \frac{\dd^3 \vb{q}}{(2\pi)^3}  \tilde{\psi}(\vb{h}'-\vb{q})  \frac{\gamma_0(\slashed{p}'+\slashed{q}+m) }{2p'\cdot q+q^2+\iu 0} \frac{e^2}{q^2-\mu_\gamma^2+\iu 0} ~,
\end{equation}
where contraction against $\bar{u}(\vb{p}')$ is implicit. 
The energy transfer is small, $q_0 \simeq (2\vb{h}'\cdot \vb{q} + \vb{q}^2)/(2 m_p)$. We will suppress the causal regulators ($+\iu 0$) in what follows. 

Let us focus on a $1s$ orbital of hydrogen for concreteness,
\begin{equation}
    \tilde{\psi}_{\rm 1s}(\vb{h}'-\vb{q}) \simeq \frac{8\sqrt{\pi}\Lambda^{5/2}}{\qty(|\vb{h}'-\vb{q}|^2 + \Lambda^2)^2}~,
\end{equation}
where we have used the non-relativistic approximation and introduced $\Lambda=1/a_0=\alpha m_e$ with $a_0$ the Bohr radius. 

It is interesting to first understand the limit of large momentum transfers. These may occur in two qualitatively different regions of phase space. First consider the final state proton's momentum to be small $|\vb{h}'|\sim \Lambda$. Then, if the Coulomb photon mediates a large momentum transfer we must have a large initial momentum for the proton which is suppressed by the hydrogenic wavefunction; in this case we find a power suppression of $\sim \Lambda^4/|\vb{q}|^{\prime 4}$. Alternatively, one can consider a large final state proton momentum with $\vb{h}'\sim \vb{q}$. In this case the integral factorizes into a tree-level matrix element (for a proton at rest) multiplied by the wavefunction at the origin $\psi_{\rm 1s}(\vb{x}=0)$ (which is suppressed by powers of $\alpha$). 

Hence the integral in \cref{1-loop} is dominated by $|\vb{q}|\sim \Lambda$ where one finds
\begin{equation}
    I^{(1)} \sim \alpha  \Lambda^{-3/2} \sim \alpha \tilde{\psi}_{\rm 1s}~,  
\end{equation}
which is $O(\alpha)$ relative to the leading expression.  We may therefore approximate $I^{(1)}$ by expanding the integrand counting $|\vb{h}'|\sim |\vb{q}|  \ll E'_e $, 
\begin{equation}\label{1-loop-atomic} 
   I^{(1)}\simeq e^2 \int  \frac{\dd^3 q}{(2\pi)^3} \ \tilde{\psi}_{1s}(\vb{h}'-\vb{q})   \frac{1 }{\vb{v}'\cdot \vb{q}} \frac{1}{\vb{q}^2+\mu_\gamma^2} ~,
\end{equation}
where $\vb{v}'=\vb{p}'/E'$; this corresponds to the eikonal approximation. To get a sense of the size of the one-Coulomb-exchange correction, we may evaluate at the kinematic point $\vb{h}'=0$. Then the integral is 
\begin{equation}
\begin{split}\label{1-loop-atomic-p0}
   I^{(1)}\rvert_{\vb{h}'=0}&\simeq e^2 \int  \frac{\dd^3 q}{(2\pi)^3} \frac{8\sqrt{\pi}\Lambda^{5/2}}{\qty(\vb{q}^2 + \Lambda^2)^2}\frac{1 }{\vb{v}'\cdot \vb{q}} \frac{1}{\vb{q}^2+\mu_\gamma^2} \\
    & = \frac{ \iu \alpha}{|\vb{v}'|}\qty[\log\qty( \frac{\mu_\gamma^2 }{\Lambda^2}) + 1] \times  \tilde{\psi}(\vb{h}'=\vb{0}) ~.
\end{split}
\end{equation}
As expected the correction is controlled by $\alpha/|\vb{v}'|\sim \alpha $ and is naively much larger than the binding corrections computed in \cite{Plestid:2024xzh}. At the special kinematic point  $\vb{h}'=\vb{0}$ the correction is purely imaginary, but is in general complex\footnote{\Cref{1-loop-atomic} has the form of a convolution and can be written in coordinate space $\int \dd^3 x  \e^{-\iu \vb{h}'\cdot\vb{x}} f(\vb{x}) \psi_{\rm 1s}(\vb{x})$. If for $\vb{h}'=\vb{0}$ this is purely imaginary, then for non-zero $\vb{h}'$ it has both real and imaginary parts.} (with non-zero real and imaginary parts) for kinematics such that $\vb{h}'\neq 0$.

If we keep the next corrections from the electron's propagator we find terms suppressed by $\Lambda/E'$ or $\Lambda/m_p$ (coming from the energy transfer $q_0=\vb{q}^2/2m_p$). As a concrete example, consider the correction proportional to $\vb{q^2}/2m_p$. Expanding  $1/(\vb{v}'\cdot \vb{q} + \vb{q}^2/2m_p)$  leads to the correction, 
\begin{equation}
    \begin{split}
    \delta I^{(1)}\rvert_{\vb{h}'=0}&\simeq \frac{e^2 }{2m_p} \int \frac{\dd^3 q}{(2\pi)^3} \frac{8\sqrt{\pi}\Lambda^{5/2}}{\qty(\vb{q}^2 + \Lambda^2)^2} \qty(\frac{1 }{\vb{v}'\cdot \vb{q}-\iu 0})^2\\
    &=\iu  \frac{\alpha}{|\vb{v}'|^2} \qty(\frac{\Lambda}{4m_p}) \times \tilde{\psi}(\vb{h}'=0)~,
    \end{split}
\end{equation}
which is suppressed by $(\Lambda/m_p)$ relative to the leading result.\!\footnote{In the case of a multi-electron atom, the final state debris can contain electrons with kinetic energy $T\sim \vb{q}^2/m_e$. In this case the suppression is $\Lambda/m_e \sim O(\alpha)$.}

\Cref{1-loop-atomic-p0} is suppressed by the explicit perturbative factor of $\alpha$, but is not suppressed by any hierarchy of scales. Coulomb corrections must therefore be controlled to at least two-loop order. As we discuss below, the Coulomb corrections actually exponentiate and cancel after integrating over the phase space of the outgoing proton. The exponentiation of Coulomb corrections can be seen transparently in coordinate space and we therefore return to the description in terms of final state wavefunctions.

\subsection{Coordinate space representation \label{Coord_Space} }
For non-zero recoil momentum, $\vb{h}'\neq 0$, we can obtain a useful representation of the integrals by noticing they have the form of a convolution. At $n^{\rm th}$-order in $(e^2)$ we encounter integrals of the form,
\begin{equation}
    \label{In_def}
    I^{(n)} =   \int \dd^3 x  ~E_{\vb{f}}^{*(n)}(\vb{x}) \e^{-\iu \vb{h}'\cdot \vb{x}} \psi(\vb{x})~. 
\end{equation}
The function $E_{\vb{f}}^{*(n)} (\vb{x})$ is a series of nested convolutions. For example $E_{\vb{f}}^{*(1)}$ is given by, 
\begin{equation}
    \label{phif_1}
    E_{\vb{f}}^{*(1)}(\vb{x})= e^{2} \int \frac{\dd^3q_1}{(2\pi)^3}\e^{\iu \vb{q}_1\cdot\vb{x}} \frac{1}{\vb{v}'\cdot \vb{q}_1} \frac{1}{|\vb{q}_1|^2} ~,
\end{equation}
and $E_{\vb{f}}^{*(2)}$ by 
\begin{align}
    \label{phif_2}
    E_{\vb{f}}^{*(2)}(\vb{x})= e^{4} \int &\frac{\dd^3q_2}{(2\pi)^3}\e^{\iu \vb{q}_2\cdot\vb{x}}\int \frac{\dd^3q_1}{(2\pi)^3}   \frac{1}{\vb{v}'\cdot \vb{q}_2}
   \\
    & \hspace{0.1\linewidth}  \times  \frac{1}{|\vb{q}_1-\vb{q}_2|^2}
    \frac{1}{\vb{v}'\cdot \vb{q}_1}\frac{1}{|\vb{q}_1|^2} ~.\nonumber
\end{align}
Inserting \cref{phif_1} and \cref{phif_2} into \cref{In_def}, writing $\psi(\vb{x}) = \int \dd^3p/(2\pi)^3 \e^{\iu \vb{p} \cdot \vb{x}} \tilde{\psi}(\vb{p})$, and carrying out the integral over $\dd^3x$ one easily arrives at the momentum space representation of $I^{(1)}$ and $I^{(2)}$ [{\it cf.}~\cref{1-loop-atomic}]. 

Next, summing over $n$ we find 
\begin{equation}
    \sum_n I^{(n)} =   \int \dd^3 x ~\psi(\vb{x}) \e^{-\iu \vb{h}'\cdot \vb{x}} E_{\vb{f}}^*(\vb{x}) ~,
\end{equation}
with 
\begin{equation}
    E_{\vb{f}}^*(\vb{x})  = \sum_n E_{\vb{f}}^{*(n)} (\vb{x})~.
\end{equation}
This resummation of diagrams is equivalent to solving the Lippmann-Schwinger equation and $E_{\vb{f}}(\vb{x})$. The function $E_{\vb{f}}(\vb{x})$ is the eikonal approximation to $\phi_{\vb{f}}(\vb{x})$. For both relativistic and non-relativistic electrons, the eikonal wavefunction satisfies \cite{Tjon:2006qe}
\begin{equation}
    \label{eikonal-diff-eq}
   \qty[ \iu \vb{v}'\cdot \vb*{\nabla} + V(\vb{x})] E_{\vb{f}}(\vb{x}) = 0~,
\end{equation}
with $V(\vb{x})$ the Coulomb potential. When approximating $\phi_{\vb{f}}$ with $E_{\vb{f}}$,  \cref{wf-norm,wf-norm-x} hold up to corrections suppressed by\footnote{In the rest frame of the atom the momentum $\vb{f}\simeq \vb{p}'$, and therefore $\vb{p}'$ labels the state.} $\Lambda/|\vb{p}'|$ as can be checked usign the method of stationary phase. The solution to \cref{eikonal-diff-eq} exponentiates in coordinate space,
\begin{equation}
    \label{eikonal-phi_f}
    E_{\vb{f}}(\vb{x}) = \exp[-\frac{\iu}{|\vb{v}'|} \int_{z}^{\infty} \dd s ~V(|\vb{x}_\perp|,s)] \equiv \e^{\iu \varphi(\vb{x})} ~,
\end{equation}
where $|\vb{x}_\perp|$ is the the spatial coordinate perpendicular to $\vb{v}'$, and $z$ labels the coordinate direction parallel to $\vb{v}'$.

\Cref{matrix-element-expression} can be written in coordinate space rather than momentum space. One then finds, 
\begin{equation}
    \mathcal{M}=\sqrt{\frac{2M_H 2 E_p}{2E_e}}\int \dd^3 x ~\e^{-\iu \vb{h}'\cdot\vb{x}} \e^{-\iu \varphi(\vb{x})}  {\sf M}(-\iu \vb*{\nabla}) \psi(\vb{x})~, 
\end{equation}
where we have abbreviated ${\sf M}(\vb{p}',\vb{k}';\vb{p},\vb{k})={\sf M}(\vb{p})$ and transformed to coordinate space where the matrix element should be interpreted as an operator ${\sf M}(-\iu \vb*{\nabla})$. 

Let us now consider the matrix element squared integrated over phase space. Importantly, in the $m_p\rightarrow \infty$ limit, the energy conserving delta function is independent of the proton's recoil momentum. Furthermore, corrections to the phase space due to a finite $\vb{h}'$ are suppressed by $\vb{h}'/E_e'$ and are negligible. We therefore have an unconstrained integral over $\vb{h}'$, 
\begin{align}
    \int \frac{\dd^3 h'}{(2\pi)^3} |\mathcal{M}|^2  = &\int  \frac{\dd^3 h'}{(2\pi)^3}  \int \dd^3x \int \dd^3 y ~\e^{\iu \vb{h'}\cdot(\vb{y}-\vb{x})} \nonumber \\
    &~~\times \e^{\iu \varphi(\vb{y})} \e^{-\iu \varphi(\vb{x})}  \\
    &\times \qty[\psi^*(\vb{y}) {\sf M}^*(-\iu \overset{\leftarrow}{\vb*{\nabla}}_y) {\sf M}(-\iu \overset{\rightarrow}{\vb*{\nabla}}_x)\psi(\vb{x}) ] ~.\nonumber
\end{align} 
The integral over $\vb{h}'$ produces $\delta^{(3)}(\vb{x}-\vb{y})$ and we find 
\begin{equation}
    \begin{split}
    \int \frac{\dd^3 h'}{(2\pi)^3} |\mathcal{M}|^2  &= \int \dd^3x 
    ~\abs{ {\sf M}(-\iu \vb*{\nabla})\psi(\vb{x})}^2\\
    &= \int \frac{\dd^3p}{(2\pi)^3} 
    ~\abs{\tilde{\psi}(\vb{p})}^2 \times \abs{ {\sf M}(\vb{p})}^2~.
    \end{split} 
\end{equation}
This leads to a result which is identical to that obtained neglecting final state interactions. We see that the Coulomb corrections cancel {\it only after} integrating over final-state phase space. This result is independent of the details of the initial state wavefunction.

\section{Multi-electron atoms \label{Multi_Electron} } 
The analysis for hydrogen is greatly simplified by the presence of a single recoiling proton in the final state. Inelastic excitations of the final state proton are gapped-out by large hadronic energy scales $\sim 300~{\rm MeV}$ and can be safely neglected. By way of contrast when considering helium and other multi-electron atoms inelastic excitations must be included. The remnant system now contains a complex spectra of excitations with energy splittings on the order of $\sim 1 {\rm Ry}= \tfrac{1}{2} \alpha^2 m_e$ and these are always energetically accessible. Furthermore, while the coupling of a proton to transverse photons is suppressed by $1/m_p$, the coupling of electrons is only suppressed by $1/m_e$, which necessitates the inclusion of both transverse and Coulomb modes. 

The addition of a single electron (i.e., going from hydrogen to helium)  therefore introduces a substantial increase in theoretical complexity. Nevertheless, as we will show below, the non-relativistic nature of the atomic debris allows us to simplify the analysis and retain theoretical control. Many of the properties derived above for hydrogen generalize readily to multi-electron atoms provided they are re-formulated at an operator level; for instance we will show that the exponentiation of Coulomb corrections persists.  In what follows we will first motivate our power counting, then analyze Coulomb photons neglecting energy splittings in the debris, and finally discuss sub-dominant corrections. 

\subsection{Power counting \label{Power_Counting} }

In this section we sketch our power counting, motivating the scaling of loop momentum, and the organizing principles behind our calculation. Following the analysis for hydrogen we take regions where $|\vb{q}|\sim \Lambda \sim \alpha m_e$ to dominate all loop integrals. This guarantees that the atomic debris remains non-relativistic in both intermediate and final states. When considering photon exchange in the non-relativistic limit it is useful to separate different modes from one another. We will phrase our power counting in Coulomb gauge which simplifies the analysis and work in the rest frame of the atom. 

Non-relativistic particles have an unsurpressed coupling to Coulomb photons. The interaction strength between two particles with charges $Q_1$ and $Q_2$ with relative velocity $\beta$ can be parameterized as $Q_1 Q_2/\beta$. For interactions among particles in the initial atom and/or atomic debris this coupling is non-perturbative, and the Coulomb interaction must be retained at leading order. We therefore include the intra-Coulomb interaction in the energies of atomic states which we will sum using completeness relations.  For Coulomb exchange with the ``struck'' electron we count Coulomb modes as $O(\alpha)$.  

The coupling of a particle to transverse photons are velocity suppressed in the non-relativistic limit. The atomic nucleus is sufficiently heavy that its velocity can be neglected entirely, while atomic electrons have velocities on the order of $v_e\sim \alpha$. We therefore count the exchange of transverse photons between the struck electron and the atomic debris as $O(\alpha v_e) \sim O(\alpha^2)$. 

Finally, in what follows we will often expand propagators making use of non-relativistic kinematics. For on-shell states $\ket{X}$ coupled to a photon with momentum $\vb{q}$,  we count energy splittings as $\Delta E_X \sim \vb{q}^2/m_e \sim \alpha |\vb{q}|$. Therefore we count powers of $\Delta E_X/|\vb{q}|$ as $O(\alpha)$. 


%
\subsection{Analysis of Coulomb photons \label{Coulomb} }
Let us first analyze the dominant final state interactions mediated by the exchange of Coulomb modes. In a multi-electron atom the wavefunction is replaced by the matrix element of the field operator with a state $\ket{B}$. Let us begin from the in-out matrix element of the weak interaction Hamiltonian density (evaluated at $x=0$), 
\begin{align}
  \mel{\nu' e' B'}{\mathcal{H}_W}{A\nu} =  \bar{u}_{k'}\gamma_\mu P_L u_k \mel{e' B'}{\bar{\psi} \Gamma^\mu \psi}{A}~.
\end{align}
Next we insert a complete set of  in-states with electric charge $Q=+e$,
\begin{equation}
    \label{break-current}
    \mel{ e' B'}{\bar{\psi}\Gamma^\mu\psi}{A} = \sumint_{B}\mel{ e' B'}{\bar{\psi}}{B}\Gamma^\mu\!\mel{B}{\psi}{A}
\end{equation}
The matrix element, $\mel{ e' B'}{\bar{\psi}}{B}$, in the above equation involves in- and out-states and can be computed using standard techniques for scattering amplitudes.

A diagrammatic interpretation of $\mel{e' B'}{\bar{\psi}}{B}$ is now clear. Drawing the highly energetic electron with a single line, and the atomic debris with a double line, we have 
\begin{equation}
\label{feynmf-pic}
\begin{gathered}
    \raisebox{10pt}{$\mel{e' B'}{\bar{\psi}}{B} ~~~=~~~~$ }
\begin{fmffile}{diagram_soft}
 \fmfstraight
\begin{fmfgraph*}(20,6.5) 
    \fmfleft{i1,i2}
    \fmfright{o1,o2}
    \fmf{double}{i1,b1,b2,o1}
    \fmf{plain}{i2,t1,t2,o2}
    \fmfv{decor.shape=square,decor.size=10,decor.filled=empty}{i2}
    \fmfpoly{square,tension=0,shaded}{b2,t2,t1,b1}
\end{fmfgraph*}
\end{fmffile}  
\raisebox{10pt}{~~~,}
\end{gathered}
\end{equation}
where the white square is the insertion of $\bar{\psi}$ and the shaded box includes all possible photon exchanges. In what follows we will expand in $\Delta E_X/|\vb{q}|$, but resumming an arbitrary number of Coulomb exchanges. 

\subsubsection*{Leading order in $\Delta E_X/|\vb{q}|$}

As shown in \cref{app-the-goods}, for Coulomb exchanges \cref{feynmf-pic} can be reduced to an operator level expression.  When $\Delta E_X$ is neglected relative to $|\vb{q}|$ the only operator that appears\footnote{If energies of intermediate states are retained then the Hamiltonian also appears in an operator-level identity .}  is the Fourier transform of the the charge density at $t=0$, $\hat{\rho}(\vb{q})$. The atomic debris is non-relativistic, the energetic electron is ballistic (i.e., eikonal), and Coulomb modes are instantaneous; taken together this implies that only one time-ordering contributes to the matrix element.   Since $\hat{\rho}(\vb{q})$, commutes with itself for different arguments, i.e., $[\hat{\rho}(\vb{q}), \hat{\rho}(\vb{q}')]=0$, the sum over orders in perturbation theory ultimately exponentiates. 

The result is that the in-out matrix element in \cref{feynmf-pic} can be written as
\begin{equation}
    \label{31-im-out-of-ideas}
    \!\!\!\mel{e'(\vb{p}') B'(\vb{h}')}{\bar{\psi}}{B(\vb{h}')} = \bar{u}(\vb{p}') \mel{B'(\vb{h}')}{\hat{W}_{\vb{v}'}}{B(\vb{h})},  
\end{equation}
where, up to $O(e^4)$, and evaluating operators at $t=0$, 
\begin{align}
    \label{W-def}
    &\hat{W}_{\vb{v}'}=1 + e^2\!\!\int \!\frac{\dd^3 q_1}{(2\pi)^3}\frac{1}{\vb{v}'\cdot \vb{q}_1} \frac{\hat{\rho}(\vb{q}_1)}{\vb{q}_1^2}  \\
    &+ e^4\!\! \int\! \frac{\dd^3 q_1}{(2\pi)^3}\frac{\dd^3 q_2}{(2\pi)^3} \frac{1}{\vb{v}'\cdot(\vb{q}_1+\vb{q}_2)} \frac{\hat{\rho}(\vb{q_2})}{\vb{q}_2^2} \frac{1}{\vb{v}'\cdot \vb{q}_1} \frac{\hat{\rho}(\vb{q_1})}{\vb{q}_1^2}~ \nonumber \\
    & + ~\ldots~ \nonumber
\end{align}
The operator $\hat{W}_{\vb{v}'}$ resembles a Wilson line, but is defined in terms of the operator $\hat{\rho}(\vb{q})$, as opposed to the photon field $\hat{A}_\mu$. Matrix elements of $\hat{W}_{\vb{v}'}$ may be thought of as resumming operator insertions of $\hat{\rho}$, 
\[
\label{wilson-line-thing}
\begin{gathered}
    \raisebox{6pt}{$ \displaystyle \langle \hat{W}_{\vb{v}} \rangle =  \mathlarger{\mathlarger{\sum_n} }\hspace{0.05\linewidth} ~$ }
\begin{fmffile}{wilson-line}
 \fmfstraight
\begin{fmfgraph*}(30,7.5) 
    \fmfstraight
    \fmfbottom{i1,b1,b2,b3,b4,b5,b6,o1}
    \fmftop{i2,t1,t2,t3,t4,t5,t6,o2}
    \fmf{phantom,tension=4}
    {i1,b1,b2,b3,b4,b5,b6,o1}
    \fmf{dbl_dots,tension=0}{b1,t1}
    \fmf{dbl_dots,tension=0}{b2,t2}
    \fmf{dbl_dots,tension=0}{b5,t5}
    \fmf{dbl_dots,tension=0}{b6,t6}
    \fmf{phantom}{t3,v,b4}
    \fmfv{label=$\ldots$\hspace{-0.045\linewidth}}{v}
    \fmf{plain,tension=4}{i2,t1,t2,t3,t4,t5,t6,o2}
    \fmfv{label=$\big\langle\hat{\rho}(\vb{q}_1\!) \hat{\rho}(\vb{q}_2) ~~\ldots~  \hat{\rho}(\vb{q}_{n-1}\!) \hat{\rho}(\vb{q}_n)\big\rangle $\hspace{-0.34\linewidth}}{b3}
\end{fmfgraph*}
\end{fmffile}  
\raisebox{6pt}{~~~~.}
\end{gathered}
\]
\bigskip

\noindent Here, the dotted lines refer to Coulomb modes (with propagators $\iu /\vb{q}^2$), and the solid line to the eikonalized electron.  See Ref.~\cite{Vaidya:2020lih} for a similar construction in the context of Glauber gluons in QCD. 

The second integral in \cref{W-def} can be symmetrized between $\vb{q}_1$ and $\vb{q}_2$. Then, using standard eikonal identities, 
\begin{equation}
    \frac12\qty(\frac{1}{\vb{v}'\cdot \vb{q}_1} + \frac{1}{\vb{v}'\cdot \vb{q}_2})= \frac12 \frac{\vb{v}'\cdot (\vb{q}_1+\vb{q}_2)}{(\vb{v}'\cdot \vb{q}_1)(\vb{v}'\cdot \vb{q}_2)}~,
\end{equation}
it is straightforward to show [in complete analogy with the eikonal wavefunction in \cref{eikonal-phi_f}] that \cref{W-def} exponentiates 
\begin{equation}
    \label{W-resum}
    \begin{split}
    \hat{W}_{\vb{v}'} &= \exp\qty[ e^2\!\!\int \!\frac{\dd^3 q}{(2\pi)^3}\frac{1}{\vb{v}'\cdot \vb{q}-\iu 0} \frac{\hat{\rho}(\vb{q})}{\vb{q}^2+\mu_\gamma^2}]\\
    &= \exp\qty[ -\iu e^2\!\!\int_{0}^\infty\dd s~ \hat{V}(\hat{s} \vb{v}')  ]~,
    \end{split}
\end{equation}
where we have made the infrared regulator $\mu_\gamma$ explicit [{\it cf.} \cref{1-loop-atomic,1-loop-atomic-p0}].  \Cref{W-resum} is the operator level generalization of \cref{eikonal-phi_f} evaluated at $\vb{x}=0$. We have introduced the Hermitian operator $\hat{V}$ 
\begin{equation}
    \hat{V}(\vb{x}) = \int \frac{\dd^3 q}{(2\pi)^3}  \e^{\iu \vb{q}\cdot \vb{x}} \frac{\hat{\rho}(\vb{q})}{\vb{q}^2+\mu_\gamma^2} ~,
\end{equation}
which manifests that $\hat{W}_{\vb{v}'}$ is unitary, $\hat{W}_{\vb{v}'}^\dagger \hat{W}_{\vb{v}'}=1$. For a point-like particle (such as the proton), the matrix element of $\hat{W}_{\vb{v}'}$ reduces to a standard Wilson line, expanded in the limit of small energy-transfer.

Upon summing over all states $\ket{B}\bra{B}$ (which is the analog of folding with the wavefunction for hydrogen), we obtain the compact formula
\begin{equation}
    \mel{e' B'}{\bar{\psi} \Gamma^\mu \psi}{A} = \bar{u}(\vb{p}') \mel{B'(\vb{h}')}{\hat{W}_{\vb{v}'} \hat{\psi}}{A(\vb{0})}~.
\end{equation}
Therefore, at the level of the cross section, we have the simple replacement \cite{Plestid:2024xzh},
\begin{equation}
    \label{W-replacement}
\begin{split}
    \!\!\mel{A}{\hat{\psi}^\dagger \delta(\ldots) \hat{\psi}}{A} 
   \rightarrow \!\mel{A}{\hat{\psi}^\dagger \hat{W}_{\vb{v}'}^\dagger \delta(\ldots) \hat{W}_{\vb{v}'} \hat{\psi}}{A}~.
\end{split}
\end{equation}
where the argument of the delta function includes the Hamiltonian, which does not commute with $\hat{W}_{\vb{v}'}$. This operator level replacement holds for arbitrary atoms. It relies only the non-relativistic dynamics of the atomic debris and on the scale separation between the energy of the outgoing electron, $E'_e$ and the binding energies of the atoms,  $\epsilon_A\sim \epsilon_B$. 

The results of this section apply to all orders in the Coulomb interaction in the limit of negligible energy transfer i.e., neglecting corrections of order $\Delta E_X/\Lambda$. Phenomenologically (e.g., for an extraction of space-like hadronic vacuum polarization) we  expect working to $O(\alpha^2)$ to be sufficient; this corresponds to two Coulomb modes. At this order, however,  sub-leading corrections not captured in \cref{W-replacement} must also be considered.

\subsubsection{Higher order in $\Delta E_X/|\vb{q}|$ \label{EX_corr}}
The leading effect arrises from corrections to the struck electron's propagator due to finite energy-level splittings, $\Delta E_X$, in the debris. The contribution from the atomic-energy $\Delta E_X$ appears at $O(\alpha \Delta E_X/|\vb{q}|) \sim O(\alpha^2)$ and must therefore also be included. Using the expansion
\begin{equation}\label{struck-e-expansion}
    {u'_0\over{u'\cdot q_1 + \iu 0}}= {-1 \over {\vb{v}' \cdot \vb{q}_1 -\iu 0}}+{q^0 \over (\vb{v}' \cdot {\vb{q}_1}-\iu0)^2} +\ldots
\end{equation}
where $u_{\mu}=\gamma(1, {\bf v'})$ is the four-velocity of the outgoing electron.

For a single Coulomb exchange (i.e., at order $e^2$) the energy transfer is given by $q_1^0=h^{'0}-h^0$. The first correction [coming from the second term on the right-hand side of \cref{struck-e-expansion}] gives rise to shift in \cref{31-im-out-of-ideas} that is proportional to 
\begin{equation}
\label{mark}
     \langle B'(\vb{h}')|\left[H,\int{\dd^3 q_1 \over (2 \pi)^3}{\hat V(\vb{q}_1) \over (\vb{v}' \cdot \vb{q}_1-\iu 0)^2}\right]|B(\vb{h}) \rangle~.
\end{equation}
Note that ${\bf q}_1$ in the integral of \cref{mark} is actually fixed by momentum conservation. The operator in the matrix element above 
\begin{equation}
   {\cal O}= \left[H,\int{\dd^3 q_1 \over (2 \pi)^3}{\hat V({\bf q}_1) \over (\vb{v}'\cdot \vb{q}_1-\iu 0)^2}\right]~,
\end{equation}
is anti-Hermitian [recall that ${\hat V}(\vb{q}_1)^{\dagger}={\hat V}(-\vb{q}_1)$] and therefore at linear order it's contribution to the cross section, which is proportional to
\begin{equation}
    \Delta \sigma \propto  \int{\dd^3 k_2 \over (2 \pi)^3}\int{\dd^3 k_1 \over (2 \pi)^3}\langle A| \hat{a}^{\dagger}_{\vb{k}_2} \left({\cal O}+{\cal O}^{\dagger} \right)\hat{a}_{\vb{k}_1} |A \rangle~,
\end{equation}
vanishes. This correction (at the amplitude level) was discussed in \ref{Pert_Theory} for the case $\bf h'=0$ when the incoming atom is hydrogen so that  $B$ and $B'$ are a single proton.

\subsection{Transverse photons \label{Trans_Photon} }
Having demonstrated the exponentiation of Coulomb modes, we now turn to transverse photons which enter first at $O(\alpha^2)$ in our power counting. Therefore, to match the size of the binding corrections considered in Ref.~\cite{Plestid:2024xzh}, we will only study the exchange of a single transverse photon.  In Coulomb gauge a crucial difference between Coulomb and transverse modes appears in their propagators:
\begin{equation}
    D_{00}(q)= \frac{\iu}{\vb{q}^2} ~~~{\rm vs.}~~ D_{ij}(q)=\frac{-\iu}{q^2+\iu 0 } \qty[ \delta_{ij}-\frac{q_i q_j}{\vb{q}^2}]~. 
\end{equation}
Transverse modes have poles in the complex $q_0$ plane and are propagating degrees of freedom. Coulomb modes are instantaneous, and non-propagating. 

When considering contributions from transverse photons it is therefore not possible to completely separate final states from initial states. We therefore organize our discussion along separate lines for final state interactions and interactions involving the initial state.

\subsubsection*{Exchange in the final state}
Let us consider a contribution to \cref{feynmf-pic} from the exchange of a single transverse photon. The spatial photon propagator in Coulomb gauge we have
\begin{equation}
    D_{ij}= \frac{-\iu}{q_0^2 - \vb{q}^2+\iu 0 } \qty[ \delta_{ij}-\frac{q_i q_j}{\vb{q}^2}] \rightarrow \frac{\iu}{\vb{q}^2-\iu 0} \qty[ \delta_{ij}]~.
\end{equation}
Where we have made use of current conservation, and an expansion in $q_0/|\vb{q}|$ in the second step. At this order (single transverse photon exchange) $q_0$ is explicitly fixed to be small by on-shell conditions from the atomic-debris side of the diagram.  

It is then clear from the structure of the loop amplitude that to this order in the expansion, transverse photons can be included by making the operator level replacement $\hat{\rho} \rightarrow \hat{\rho} - \vb{v}'\cdot \hat{\vb{J}}$. Counting non-relativistic matrix elements of the current operator $\langle \hat{\vb{J}} \rangle \sim O(\alpha)$  then guarantees that transverse photons do not spoil the exponentiation at $O(\alpha^2)$. 

\subsubsection{Exchange with the initial state}
There are also corrections that arise from the exchange of photons between the outgoing struck electron and the initial atom. Hard photons, with virtuality on the order of $m_e^2$ or larger, will factorize from bound state effects and are already included in existing calculations of QED corrections to $\nu e \rightarrow \nu e$ scattering with free electrons. 

Corrections involving soft photons may, however, be sensitive to atomic structure. Any transverse photon coupling from the atom to the debris is doubly suppressed by non-relativistic velocities and enters at $O(\alpha^3)$. We therefore focus on diagrams (shown below) in which the transverse photon couples to the outgoing ballistic electron. It is instructive to compare the one-loop diagrams involving the exchange of a Coulomb (dotted line) vs.~a transverse (wavy line) photon,  
\[
\begin{gathered}
\begin{fmffile}{coulomb-graph}
 \fmfstraight
\begin{fmfgraph*}(16,16)
    \fmfstraight
    \fmftop{t}
    \fmfleft{ib,i1,i2,i3,i4,i5,it}
    \fmfright{ob,o1,o2,o3,o4,o5,ot}
    \fmfbottom{b}
    \fmf{dashes}{b,t}
    \fmf{double,tension=2}{i3,v1,v_hard}
    \fmf{phantom}{v_hard,o3}
    \fmffreeze
    \fmf{double}{v_hard,o1}
    \fmf{plain,tension=1.1}{v_hard,v2,ot}
    \fmf{dbl_dots,tension=0,left=0.5,label=$q$}{v1,v2}
    \fmfv{decor.shape=square,decor.size=10,decor.filled=empty}{v_hard}
    \fmfv{label=$A$}{i3}
    \fmfv{label=$B$}{o1}
    \fmfv{label=$e$}{ot}
\end{fmfgraph*}
\end{fmffile}  
\hspace{0.2\linewidth}
\begin{fmffile}{transverse-graph}
 \fmfstraight
\begin{fmfgraph*}(16,16)
    \fmfstraight
    \fmftop{t}
    \fmfleft{ib,i1,i2,i3,i4,i5,it}
    \fmfright{ob,o1,o2,o3,o4,o5,ot}
    \fmfbottom{b}
    \fmf{dashes}{b,t}
    \fmf{double,tension=2}{i3,v1,v_hard}
    \fmf{phantom}{v_hard,o3}
    \fmffreeze
    \fmf{double}{v_hard,o1}
    \fmf{plain,tension=1.1}{v_hard,v2,ot}
    \fmf{photon,tension=0,left=0.5,label=$q$}{v1,v2}
    \fmfv{decor.shape=square,decor.size=10,decor.filled=empty}{v_hard}
    \fmfv{label=$A$}{i3}
    \fmfv{label=$B$}{o1}
    \fmfv{label=$e$}{ot}
\end{fmfgraph*}
\end{fmffile}  
\raisebox{20pt}{~~~.}
\end{gathered}
\]
We have drawn a dashed line where we insert a complete set of states in \cref{break-current}. 
The Coulomb graph vanishes, because its propagator supplies no poles in the complex $q_0$ plane, and the poles from the atom and the debris both lie on either the upper- or lower-half of the complex plane (see Appendix B of \cite{Plestid:2024eib} for a more detailed discussion). A similar argument applies to Coulomb exchanges between the initial state atom and final state debris. Physically these results follow from the non-relativistic nature of the bound-state constituents, and the fact that the Coulomb interaction is instantaneous. 

The graph involving a transverse photon has additional poles from the photon propagator that invalidate the argument given above. Performing the integral over $q_0$ by closing  the contour in the upper half plane picks up {\it only} the photon's pole. This corresponds to an intermediate state that contains a photon [and contributes to the sum in \cref{break-current}]. 
{However transverse photons couple proportional to a three momentum (i.e., $\hat{\vb{J}} = \sum_i \hat{\vb{p}}_{e,i}/m_e$)  and in the scattering amplitude this is integrated against the initial state multiparticle wave function.  For rotationally invariant targets this integral gives a contribution to the cross section that vanishes,\!\footnote{Strictly speaking, the expectation value of momentum can be proportional to the spin of the atom. For unpolarized targets this cancels upon spin averaging. } because $\langle \hat{\vb{p}}_e \rangle \simeq0$. Therefore, the contribution from transverse photons to the cross section can occur first at order $e^2 \langle \hat{\vb{p}}^2 \rangle /m_e^2 \sim O(\alpha^3)$.}



\section{Applications \label{Applications}} 
%
We now return to the phenomenological examples discussed in the introduction. We begin with neutrinos scattering from atomic electrons. This is convenient since in the previous sections we have already analyzed $\nu e \rightarrow \nu e$ scattering for atomic electrons in detail. In the absence of final-state interactions, the cross section\footnote{The same formula applies for the differential cross section $\dd \sigma /\dd \Pi_e$ where $\Pi_e$ is the phase space of the struck electron \cite{Plestid:2024xzh}.} for neutrino-atom scattering can be written as \cite{Plestid:2024xzh}, 
\begin{align}
   \sigma &=  \int \dd \epsilon \int \frac{\dd^3 p}{(2\pi)^3 } S_A(\epsilon,\vb{p})\\
   &\hspace{0.15\linewidth}\frac{1}{2 \omega} \frac{1}{2E}\int \dd \Pi_e \frac{1}{2\omega'} \delta({\cal E}-\epsilon)\frac12\sum_{\rm spins} |{\sf M}|^2 ~.\nonumber
\end{align}
 Here ${\sf M}(\vb{k}',\vb{p}';\vb{k},\vb{p})$, is the free-particle matrix element for  $\nu(\vb{k}) e(\vb{p}) \rightarrow \nu(\vb{k}') e(\vb{p}')$ scattering and $S_A(\epsilon,\vb{p})$ is the non-relativistic spectral function of the target atom $A$, 
\begin{equation}
    S_A=\!\!\int \!\frac{\dd^3 q}{(2\pi)^3}\mel{A}{\hat{a}^\dagger_{\vb{q}}  \delta(\epsilon_A+\hat{H}-\epsilon)  \hat{a}_{\vb{p}} }{A}~.
\end{equation}
Binding corrections can be efficiently incorporated by expanding $|{\sf M}|^2$ in $\epsilon$ and $\vb{p}$~\cite{Plestid:2024xzh}. 

To include final-state Coulomb exchange one simply makes the replacement $S_A \rightarrow \tilde{S}_A$ where 
\begin{equation}
    \label{tilde-S}
    \tilde{S}_A=\!\!\int \!\frac{\dd^3 q}{(2\pi)^3}\mel{A}{\hat{a}^\dagger_{\vb{q}} \hat{W}_{\vb{v}'}^\dagger \delta(\epsilon_A+\hat{H}-\epsilon) \hat{W}_{\vb{v}'} \hat{a}_{\vb{p}} }{A}~.
\end{equation}
The same procedure (mentioned above) of expanding $|{\sf M}|$ in terms of $\vb{p}$ and $\epsilon$ still applies. We note that \cref{tilde-S} is reminiscent of the opperatorial definition of a parton distribution function.

The unitarity of $\hat{W}_{\vb{v}'}$ guarantees that corrections disappear after integrating over $\dd \epsilon$ (which removes the delta function). The first term for which this is not true is the energy weighted sum $\int \dd \epsilon (-\epsilon) \times (\ldots)$, which introduces the insertion of the Hamiltonian. The non-commutativity of $\hat{H}$ and $\hat{W}_{\vb{v}'}$ then gives
\begin{align}
    \int \dd \epsilon \tilde{S}_A(\epsilon,\vb{p}) &=  \int \!\frac{\dd^3 q}{(2\pi)^3}\mel{A}{\hat{a}^\dagger_{\vb{q}}   \hat{a}_{\vb{p}} }{A}~,\\
    \int \dd \epsilon (-\epsilon) \tilde{S}_A(\epsilon,\vb{p}) &=  \!\!\int \!\frac{\dd^3 q}{(2\pi)^3}\mel{A}{\hat{a}^\dagger_{\vb{q}} \hat{W}_{\vb{v}'}^\dagger  [\hat{W}_{\vb{v}'}\hat{a}_{\vb{p}},\hat{H}] }{A}~.
    \label{mod-Koltun}
\end{align}
\Cref{mod-Koltun} generalizes the Koltun sum rule \cite{Koltun:1972kh} to include final-state Coulomb interactions in the leading non-relativistic approximation. 

In our counting $\epsilon\sim O(\alpha^2)$, and so when computing $O(\alpha^2)$ corrections to the cross section we can set $\hat{W}_{\vb{v}'}=1$ in \cref{mod-Koltun}. This demonstrates that Coulomb corrections do not alter the predictions of the cross section until at least $O(\alpha^3)$. 

For the same reason, when considering the sub-leading $O(\alpha^2)$ corrections we only need to interfere these contributions with the tree-level matrix element evaluated at $\vb{p}=0$ and $\epsilon=0$ (hereafter referred to as ``1''). 
As argued in \cref{Coulomb,Trans_Photon} the contributions from final state transverse photon exchange, and corrections due to finite energy splitting in the debris do not contribute to the cross section at $O(\alpha^2)$.  We therefore conclude that all final-state interaction effects can be neglected at $O(\alpha^2)$ for observables that are inclusive with respect to the atomic debris (i.e., that have integrated over the debris' phase space). We note however, that corrections involving transverse photon exchange with the initial atom must still be evaluated for a complete $O(\alpha^2)$ analysis. 

\subsection{Neutrino flux normalization}
One application of our results is for neutrino flux normalizations at long-baseline experiments such as DUNE and T2HK.  These experiments require the $\nu e \rightarrow \nu e$ cross section to be known with a precision on the order of (ideally at least a factor of ten smaller than) one percent. In reality these experiments measure hard outgoing electrons from atomic scattering. For example in the case of DUNE the relevant process is $\nu {\rm Ar} \rightarrow \nu e B$ with $B$ all the soft atomic debris. 

As discussed in \cite{Plestid:2024xzh}, atomic binding corrections are much smaller than power-counting would suggest due to accidental cancellations that come from the structure of the tree-level matrix element. Coulomb corrections vanish upon integrating over the final state debris, and transverse photon exchange vanishes at least until $O(\alpha^2)$.   We therefore conclude that all possible sources of QED corrections to $\nu A \rightarrow \nu e B$ scattering are under sufficient control (i.e., better than $1\%$) for neutrino flux normalization.

\subsection{Extraction of hadronic vacuum polarization}
Another process for which the results of this paper are relevant is the $t$-channel scattering of charged leptons off atomic electrons. We have in mind (in particular) muons scattering on electrons bound to carbon 
\begin{equation}
    \mu(\vb{k}) \!~^{12}{\rm C}(\vb{0}) \rightarrow \mu(\vb{k}') e(\vb{p}') B(\vb{h}')~.
\end{equation}
This is relevant for the MuonE experiment \cite{Abbiendi:2016xup,Abbiendi:2022oks} which aims to extract hadronic vacuum polarization at space-like kinematics. MuonE's data can potentially give an independent determination of hadronic vacuum polarization's contribution to the anomalous magnetic moment of the muon \cite{CarloniCalame:2015obs,Abbiendi:2016xup,Abbiendi:2022oks}. The theory demands of the experiment are a relative precision of $10^{-5}$ \cite{Banerjee:2020tdt}, which necessitates the inclusion of bound state effects. 

In Ref.~\cite{Plestid:2024xzh} we have computed the shift in the shape of $\dd \sigma /\dd t$ due to atomic binding effects. The results of this paper imply that final-state photon exchange between the struck electron and the atomic debris do not alter the conclusion of our previous work at $O(\alpha^2)$ (the working-order of Ref.~\cite{Plestid:2024xzh}). 

Muon electron scattering, however, involves a charged muon which may also exchange photons with both the initial atom and the outgoing debris. The results of this paper are therefore not sufficient for a complete account of all of these effects, however we can comment immediately on a sub-class of diagrams. Let us first match the hard-photon mediated $\mu-e$ scattering vertex onto an effective four-Fermi contact operator, $\hat{\mathcal{O}}$. The operator's Wilson coefficient includes all short-distance radiative corrections.  

We are then interested in matrix elements of the form $\mel{B'e'\mu'}{\hat{\mathcal{O}}}{A\mu}$. If we insert a complete set of states before the operator insertion, we can re-sum Coulomb exchange into an operator level expression involving $\hat{W}_{\vb{v}_\mu}^\dagger$ where $\vb{v}_{\mu}$ is the 3-velocity of the initial muon. We can resum Coulomb exchange between the final state leptons and the atomic debris as described above using the product of $\hat{W}_{\vb{v}_\mu'}$ and   $\hat{W}_{\vb{v}_e'}$. These operators commute with one-another since they only involve $\hat{\rho}(\vb{x})$ or equivalently $\hat{\rho}(\vb{q})$. Our experience with final state Coulomb exchanges suggests that for high energy muon atomic electron scattering, all the effects of soft Coulomb exchanges can be captured by making the replacement  
\begin{equation}
    \label{conjecture}
    \hat{a}_{\vb{p}} \rightarrow \hat{W}_{\vb{v}_\mu'}\hat{W}_{\vb{v}_e'} \hat{a}_{\vb{p}} \hat{W}_{\vb{v}_\mu}^\dagger~,
\end{equation}
in the definition of the spectral function. The physical interpretation being that the Wilson-line before $\hat{a}_{\vb{p}}$ ``shakes up'' the initial state. This leads to different sum rules than in the case of neutrino electron scattering, 
\begin{equation}
    \int \dd \epsilon  \tilde{S}_A(\epsilon,p) =  \int \!\frac{\dd^3 q}{(2\pi)^3}\mel{A}{\hat{W}_{\vb{v}_\mu}\hat{a}^\dagger_{\vb{q}}   \hat{a}_{\vb{p}} \hat{W}_{\vb{v}_\mu}^\dagger}{A}~.
\end{equation}
We are interested in whether or not the operators $\hat{W}_{\vb{v}_\mu}$ and $\hat{W}_{\vb{v}_\mu}^\dagger$ modify the sum rules that are relevant for our analysis. Corrections to the Koltun sum rule, and to the average kinetic energy are necessarily $O(\alpha^3)$ and can therefore be neglected in our $O(\alpha^2)$ analysis.

The leading-order sum rule, $\int \dd^3p/(2\pi)^3 \int \dd \epsilon \tilde{S}_A(\epsilon,p) = Z$, could in principle receive an $O(\alpha)$ or $O(\alpha^2)$ correction. Integrating over $\dd^3 p$, using $\hat{\psi}(0)= \int \dd^3 p/ (2\pi)^3 \hat{a}_{\vb{p}}$ and carrying out both integration over $p$ and $q$ we find
\begin{equation}
    \int \frac{\dd^3 p}{(2\pi)^3}  \int \dd \epsilon  \tilde{S}_A(\epsilon,p) = \mel{A}{\hat{W}_{\vb{v}_\mu}\hat{\rho}_e \hat{W}_{\vb{v}_\mu}^\dagger}{A}~.
\end{equation}
where $\hat{\rho}_e=\hat{\psi}^\dagger \psi$ and the fields are evaluated at $x=0$. Then, since the electron charge density $\hat{\rho}_e$ commutes with $\hat{W}_{\vb{v}_\mu}$ we conclude that (provided \cref{conjecture} holds) the modified spectral function satisfies the same sum rule as when Coulomb exchanges are ignored, 
\begin{equation}
    \!\!\!\int \!\frac{\dd^3 p}{(2\pi)^3} \!\! \int \dd \epsilon  \tilde{S}_A(\epsilon,p) = \!\!\int\! \frac{\dd^3 p}{(2\pi)^3}\!\!  \int \dd \epsilon  S_A(\epsilon,p) = Z ~,
\end{equation}
where $Z$ is the atomic number of the state $A$. This is not surprising since Coulomb exchanges do not change the number of electrons in the atom. We therefore conclude that Coulomb exchanges in both the initial and final state do not interfere with the leading order amplitude for $\mu \!~^{12}{\rm C} \rightarrow \mu e B^+$ scattering until at least $O(\alpha^3)$. 


\section{Conclusions \label{Conclusions} } 

We have considered corrections to scattering cross sections that arise when a target electron is bound in an atom. The focus of the present paper is on final-state photon exchange and compliments our previous work on corrections due to the initial bound state \cite{Plestid:2024xzh}. 

We have studied which regions of loop momentum dominate virtual corrections, and find that photons with virtuality on the order of the atomic scale $\Lambda \sim \alpha m_e$ dominate the integrals. We have organized these virtual corrections using a power counting scheme that makes use of the non-relativistic features of the atomic bound state, $A$, and final state debris, $B^+$. 

Within this power-counting, Coulomb exchanges provide the dominant leading correction (before summing over final states). We have studied these graphs in detail and find that they can be resummed in the leading non-relativistic approximation at an operator level. This includes arbitrary multi-particle intermediate states (e.g., $^{12}{\rm C}^{3+} e^-e^-$) with charge $Q=+1$.  Our work then provides an operator level definition of Coulomb corrections in the eikonal approximation for a system with small excitation energies. We find that the unitary nature of the resummed Coulomb corrections (they resemble a Wilson line) demonstrates that these corrections do not enter until at least $O(\alpha^3)$ for neutrino electron scattering. Our experience suggests that they likely do not enter for $\mu e \rightarrow \mu e$ scattering either (this second result relies on \cref{conjecture} holding). 

We have also considered sub-leading corrections in the final state, such as transverse photon exchange (suppressed by non-relativistic velocities in the atom) and corrections to Coulomb exchanges from finite-energy splittings among different final states in the atomic debris. We find that both of these corrections (already down by $\alpha^2$) do not interfere with the tree-level cross section after integrating over the final states of the debris, and therefore do not influence the cross section at $O(\alpha^2)$. We conclude that all effects necessary for a sub-percent prediction of the neutrino (atomic) electron cross section are under control.

In the context of $\mu e \rightarrow \mu e$ scattering, the results of Ref.~\cite{Plestid:2024xzh} and of the current work account for all of the leading ``$Z$-enhanced'' corrections to the cross section. These correspond to the class of diagrams that are $O(Q_e^2 Q_B^2)$, with $Z$-enhancements in medium-heavy atoms; the correction to the cross section per electron scale parametrically as $\sim Z^{4/3} \alpha^2$. 

\pagebreak 

At $O(e^4)$, there can be further corrections from final state interactions that stem from ``mixed'' corrections involving both Coulomb and transverse photons e.g., 
\bigskip
\begin{equation*}
\nonumber
\begin{gathered}
\begin{fmffile}{diagram_mixed}
 \fmfstraight
\begin{fmfgraph*}(20,6.5) 
    \fmfstraight
    \fmfleft{i1,i2}
    \fmfright{o1,o2}
    \fmf{double}{i1,b1,o1}
    \fmf{plain}{i2,t1,t2,t3,o2}
    \fmf{photon,tension=0,left=0.5}{t1,t3} 
    \fmf{dbl_dots,tension=0}{t2,b1}
    \fmfv{decor.shape=square,decor.size=10,decor.filled=empty}{i2}
\end{fmfgraph*}
\end{fmffile}  
\raisebox{10pt}{~~~.}
\end{gathered}
\end{equation*}
These only receive $Z$-enhancements for Coulomb mode, but not for the transverse photon. Furthermore, loops with a transverse photon scale like $\alpha/(4\pi)$, whereas Coulomb modes scale like $\alpha/|\vb{v}'| \sim \alpha$. We therefore expect that these ``mixed'' corrections scale as $\sim Z^{2/3}\alpha^2/(4\pi)$ which is smaller than the purely kinematic binding corrections by\footnote{This estimate does not account for any large logarithms (or accidental numerical enhancements) that may appear in the mixed radiative corrections. } $Z^{-2/3}/(4\pi) \approx 0.02$. We therefore defer their evaluation to future work.


In summary, we have made further progress on understanding radiative corrections to leptonic interactions with atomic electrons. We find that photon exchanges (in both the initial and final state) between the atomic system and any high energy leptons cancel after integrating over the phase space of the atomic debris. The formalism for Coulomb corrections developed in this work is appropriate for systems with many low-lying energy levels that should be summed over. This is different from conventional treatments in terms of a static background potential. The results in this paper account for all maximally $Z$-enhanced corrections at $O(\alpha^2)$. The standard non-relativistic power-counting applied in this work should serve as a useful guide for future studies of atomic corrections to muon electron scattering for the MuonE experiment.

\section*{Acknowledgements}
Some of this work was done at the Aspen Center for Physics, which is supported by National Science Foundation grant PHY-1607611.  RP is supported by the Neutrino Theory Network under Award Number DEAC02-07CH11359. RP and MW are supported by the U.S. Department of Energy, Office of Science, Office of High Energy Physics under Award Number DE-SC0011632, and by the Walter Burke Institute for Theoretical Physics.

\pagebreak

\appendix

\section{Exponentiation of Coulomb corrections \label{app-the-goods} }
In this section we derive \cref{W-def} by considering the perturbative series that arises from Coulomb exchange. We show that within a non-relativistic approximation, the Coulomb exchange leads to an expression involving a unitary operator that is closely related to a Wilson line . The main novelty of our derivation is the inclusion of arbitrary states of the atomic debris between subsequent insertions of the charge-density operator. In particular, we do not model Coulomb exchanges using a (classical) background potential, and instead include all possible non-relativistic excited states of the debris as intermediate states in the perturbative expansion. 

We find that only one time-ordering contributes (the uncrossed ladder graphs). After expanding the ballistic electron propagator to take advantage of the small energy transfers (fixed by the non-relativistic atomic debris) the expressions are independent of intermediate state energies at leading order. This allows the intermediate states to be removed via completeness, and a simple operator-level identity remains. We now discuss two different derivations, on in the time-domain and one in the frequency-domain. 

\subsection{Marks Way \label{app-Marks-way} }

It is convenient to define,
\begin{equation}
    \label{Marks_eq1}
    \int \dd^4x \e^{\iu Q\cdot x}\langle B'(\vb{h}')e(\vb{p}')|\bar{\psi}(x)|B(\vb{h}) \rangle = {\bar u}I~,
\end{equation}
where $Q$ is a residual electron four momentum and our convention is that the final state electron has zero residual momentum. Factoring out the energy momentum conserving delta function we write, $I(q_1)=(2\pi)^4\delta({Q}+(h'-h))J(q_1)$. Next, $I$ (or $J$) can be expanded in a series $I=I^{(0)}+I^{(1)}+ \ldots$  where the superscript is the number of Coulomb exchanges. The leading order piece is $I^{(0)}=(2\pi)^4\delta({Q}+(h'-h))$, or equivalently $J^{(0)}=1$ and for the one photon exchange in Coulomb gauge $I^{(1)}$ we have, treating the electron as ballistic and using the appropriate vertex and propagator,
\begin{equation}
    I^{(1)}=e^2{ u^{\prime0} \over u'\cdot Q}\int \dd^4x_1 \e^{\iu Q\cdot x_1} \langle B'(\vb{h}')|J_0(x_1)|B(\vb{h})\rangle ~.
\end{equation}
Here $u^{\mu}$ is the electron four velocity ($u^{\mu}=\gamma(1, {\bf v'})$ in the rest frame of the atom) and we have used the fact that the constituents of the states are non-relativistic so matrix elements of the electromagnetic 3-vector current are suppressed. Translating the charge density operator $J_0(x)\rightarrow J_0(x=0)$ to the origin of space-time and performing the integration over $ x_1$ the above is equal to 
\begin{align}
    I^{(1)}={ u^{\prime0}\over u'\cdot Q} {e^2 \over {\bf Q}^2} (2\pi)^4&\delta(Q+(h'-h))\\
    &\langle B'(\vb{h}')|J_0(0)|B(\vb{h})\rangle~. \nonumber 
\end{align}
Next we use 
\begin{equation}
  J_0(0)=\int {\dd^3q_1 \over (2\pi)^3} {\tilde J}_0({\bf q}_1) ~, \vspace{6pt}
\end{equation}
and  $\langle B'(\vb{h}')|{\tilde J}_0({\bf q})|B(\vb{h})\rangle \propto (2\pi)^3\delta({\bf Q}-{\bf q})$. Note that ${\tilde J}_0({\bf q}_1)$ is also evaluated at $t=0$. It can be combined with the factor $1/{\bf q}_1^2$ into the Fourier transform of the potential energy operator ${\tilde V}$. Finally since non relativistic energies  are small compared to momenta we can write $u'\cdot Q\simeq -u'^0{\bf v'}\cdot {\bf Q}$ yielding
\begin{equation}
    J^{(1)}= e^2 \langle B'(\vb{h}')|\int {\dd^3q_1 \over (2\pi)^3}{{\tilde V({\bf q}_1})\over{\bf v'}\cdot {\bf q}_1}|B(\vb{h})\rangle~.
\end{equation}
At second order we have
\begin{align}
& \!\!\!\!  I^{(2)}= \int{\dd^4 q_2 \over (2 \pi)^4}{ u^{\prime0}\over u'\cdot q_2+\iu 0}{ u^{\prime0}\over u'\cdot Q +\iu0}{ e^2 \over {\bf q}_2^2} { e^2 \over |{\bf Q}-{\bf q}_2|^2} ~~~~~~\nonumber\\
& \hspace{0.125\linewidth}  \int{\dd^4 x}\!\int{\dd^4 y} ~\e^{-\iu (Q-q_2)\cdot x}\e^{-\iu q_2\cdot y} \\
&\hspace{0.2\linewidth}\langle B'(\vb{h}')|T\{J_0(y)J_0(x)\}|B(\vb{h})  \rangle ~.\nonumber  
\end{align} 
First imagine doing the $q_2^0$ integration (see \cref{app-Ryans-way} for an explicit discussion of the method by contours). The ballistic electron propagator only has a pole in the lower half $q_2^0$ complex plane. Consequently in the time ordered product of the currents only the time ordering where $y^0-x^0 >0$ contributes and we can so we can replace\footnote{Since $1=\theta(x^0-y^0)+\theta(y^0-x^0)$ and the second theta function gives no contribution.} 
\begin{equation*}
    T\{J_0(x)J_0(y)\}\rightarrow J_0(y)J_0(x)\theta(y^0-x^0)\rightarrow J_0(y)J_0(x)~,
\end{equation*}
in the integral. Next insert a complete set of states $|X_1(p_{X_1})\rangle$ between the two currents and translate the currents to the origin of space-time.
Then perform the coordinate integrations. This gives  a product of two delta functions, $(2\pi)^4\delta(p_{X_1}-h'+q_2)(2\pi)^4\delta(h-p_{X_1}-q_2+Q)$. They  can be rewritten as a product of  the overall four momentum conservation delta function and one that performs the $q_2$ integration setting $q_2=h'-p_{X_1}$.  This yields
\begin{widetext}
\begin{equation}
    J^{(2)} \simeq  \sumint_{X_1} {1\over {\bf v'}\cdot({\bf h'}-{\bf h})}{1 \over {\bf v'}\cdot({\bf h'}-{\bf p}_{X_1})}
    {   e^2 \over |{\bf h'}-{\bf p}_{X_1}|^2} {e^2 \over |{\bf h}-{\bf p}_{X_1}|^2}
    \langle B'(\vb{h}') | J_0(0)|X_1(p_{X_1})\rangle \langle X_1(p_{X_1})|J_0(0)|B(\vb{h})\rangle~.
\end{equation}

Next express the charge density operator at the origin and an integral over the three dimensional Fourier transform, and use the fact that matrix elements conserve momentum to reassemble terms into the Fourier transform of the potential operator $\tilde V$. At this point the sum over intermediate states can be done yielding
\begin{equation}
    J^{(2)} \simeq e^4 {1 \over {\bf v'}\cdot ({\bf h}'- {\bf h})} \langle B'(\vb{h}')|\int {\dd^3 q_1 \over (2\pi)^3 }{\tilde V}({\bf q}_2){1 \over {\bf v'}\cdot {\bf {q}_2}} \int {\dd^3 q_1 \over (2\pi)^3 }\tilde{V}(\vb{q}_1) |B(\vb{h})\rangle ~.
\end{equation}
\end{widetext}
Since the matrix element under the integral is proportional to $\delta^{(3)}([\vb{h}-\vb{h}']-[\vb{q}_1+\vb{q}_2] )$ this can be rewritten as, {\it cf}.~\cref{W-def}, 
\begin{align}
    J^{(2)}=  e^4 \int {\dd^3 q_1 \over (2\pi)^3 }& \int  {\dd^3 q_2 \over (2\pi)^3 } {1 \over {\bf v}' \cdot ({\vb{q}_1} +\vb{q_2})} {1 \over {\bf v'}\cdot \vb{ q}_2} \\
  &\langle B'(\vb{h}')|{\tilde V}({\bf q}_2) V({\bf q}_1) |B(\vb{h})\rangle ~.\nonumber
\end{align}
Obviously this procedure can be repeated to deduce expressions for all the $J^{(n)}$.

\subsection{Ryan's way \label{app-Ryans-way} }
Let us repeat the above analysis with a different emphasis and method. Consider the $n^{\rm th}$-order in perturbation theory. Photon exchange between the electron line and the charged-debris $\ket{B}$ leads to a conventional diagrammatic expansion (with fermion and photon propagators) and a ``blob'' on the bottom of the diagram with an incoming ionic system $B$ and an outgoing system $B'$. This ``blob'' corresponds to a multi-current correlator, $G(q_1,\ldots, q_n)$ of time-ordered electromagnetic currents (specifically $\hat{J}_0=\hat{\rho}$ for Coulomb exchange), 
\begin{equation}
    \label{blob-def}
   G^{(n)}=\!\!\int \!\!\{ \dd x\} \e^{+\iu \sum q_i \cdot x_i}\!  \mel{B'}{T\{ \hat{J}_{0}(x_1) \ldots \hat{J}_{0}(x_n)\}}{B}.
\end{equation}
In analogy with the transition from $I$ to $J$ discussed above, it is convenient to factor out an overall momentum conserving delta function and introduce $\mathcal{G}$. 
\begin{equation}
    \begin{split}
    G^{(n)}(q_1,\ldots, q_n) = &(2\pi)^4 \delta(\Sigma q- Q) \\
    &~~\times \mathcal{G}^{(n)}(Q;~q_2,\ldots, q_N) ~.
    \end{split}
\end{equation}
where $Q = h'-h$  is enforced by the delta function. Notice that $\mathcal{G}^{(n)}$ is a function of $n-1$ variables, while $Q$ is a parameter fixed by the states $\ket{B(\vb{h})}$ and $\bra{B'(\vb{h}'}$. 
Making use of the eikonal approximation on the highly energetic electron leg, and working in Coulomb gauge, we then find
\begin{widetext}
\begin{equation}
    \mel{e' B'}{\bar{\psi}}{B} = \bar{u}(\vb{p}') \qty[ 
\braket{B'}{B} + e^2\frac{1}{\vb{q}_1^2} \frac{u'_0}{u'\cdot \vb{Q}}  \mathcal{G}^{(1)}(Q) + e^4\int\frac{\dd^4 q_2}{(2\pi)^4}  \frac{1}{\vb{q}_2^2} \frac{u'_0}{u'\cdot q_2}  \frac{1}{|\vb{Q}-\vb{q}_2|^2} \frac{u'_0}{u'\cdot  Q} \mathcal{G}^{(2)}(Q;~q_2) + \ldots] ~.
    \label{PT_with_G}
\end{equation}
Extension to higher orders is straightforward: one appends an extra integral, an extra fermion and photon propagator, and uses $\mathcal{G}^{(n)}(Q;~q_2,\ldots, q_n)$. The term in square brackets in \cref{PT_with_G} corresponds to $J$ as defined beneath \cref{Marks_eq1}. 

It is useful to write $\hat{J}_0(0)$ in terms of its components which supply definite momentum transfer $\vb{k}$. For example,
\begin{equation}
    \begin{split}
    \mathcal{G}^{(1)}(Q)= \mel{B'(\vb{h}')}{\hat{J}_0(0)}{B(\vb{h})} = \int \frac{\dd^3 \vb{q}_1}{(2\pi)^3} \mel{B'}{\hat{\rho}(\vb{q}_1)}{B} ~,
    \end{split}
\end{equation}
where $\hat{\rho}(\vb{q}_1) = \int \dd^3x ~\e^{-\iu \vb{q}_1\cdot \vb{x}}\hat{J}_0(\vb{x})$. 

We now proceed with a direct evaluation of the loop integrals by contour integration. Let us study $\mathcal{G}^{(2)}(Q;~q_2)$ as an explicit example. Begin by explicitly time ordering the currents in $\mathcal{G}^{(2)}$. Inserting a complete set of states, re-writing $\hat{J}_0(0)$ in terms of $\hat{\rho}(\vb{q}_1)$, and carrying out the remaining integral over $\dd^4x$ we obtain (defining $\omega_2=q_2^0$), 
\begin{equation}
    \label{expl-time-order}
    \begin{split}
    \!\!\mathcal{G}^{(2)}(Q;~q_2)= \!\!\!\int \!\!\frac{\dd^3q_1}{(2\pi)^3}\!\sumint_X \!\mel{B'}{\hat{\rho}(\vb{q}_1)}{X} \frac{-\iu}{\omega_2-\Delta E_X +\iu 0} \mel{X}{\hat{\rho}(\vb{q}_2)}{B}
    \!+\!
    \mel{B'}{\hat{\rho}(\vb{q}_2)}{X} \frac{-\iu}{\omega_2 -\Delta E_X'-\iu 0 } \mel{X}{\hat{\rho}(\vb{q}_1)}{B},
    \end{split}
\end{equation}
\end{widetext}
where $\Delta E_X = E_X - E_B$ and $\Delta E_X'=E_X-E_B'$. This result is obtained by translating the currents to $t=0$ and integrating the exponentials. It essentially amounts to a derivation of time-ordered perturbation theory, and the two terms in \cref{expl-time-order} can be interpreted as a crossed and uncrossed Feynman diagram.

Next notice that because $1/u'\cdot q_2 \equiv 1/(u'\cdot q_2 +\iu 0)$ only has a pole in the lower-half plane. We may then close our contour in the upper-half plane which selects only the second term in \cref{expl-time-order}. At higher orders in perturbation theory all time orderings with poles in the lower-half plane vanish after contour integration; these correspond to crossed ladders. It follows that only the uncrossed ladders contribute to the amplitude to any order in perturbation theory, in agreement with the time-domain analysis presented in \cref{app-Marks-way}.

Expanding the eikonal propagators and dropping the (now on-shell) energy transfer $\Delta E_X$ gives, 
\begin{equation}
    \qty[\frac{u'_0}{u'\cdot q_2 +\iu 0}]_{\omega_2=\Delta E_X'} \simeq \frac{1}{-\vb{v}'\cdot \vb{q}_2 +\iu 0} \left(1+O\qty(\frac{\Delta E_X'}{|\vb{q}|})\right)~. 
\end{equation}
The propagator structure is then independent of $\Delta E_X$, and the sums over intermediate states reduce to the identity by completeness. We are then left with \cref{W-def}, with the $n^{\rm th}$ order contribution being given by 
\begin{widetext}
\begin{equation}
    e^{2n} \int\qty[\frac{\dd^3q_1}{(2\pi)^3} \ldots \frac{\dd^3q_n}{(2\pi)^3} ]\qty{\frac{1}{\vb{v}'\cdot\vb{q}_1}\frac{1}{\vb{v}'\cdot(\vb{q}_1+\vb{q}_2)} \ldots \frac{1}{\vb{v}'\cdot (\sum_i \vb{q}_i)}} \prod_i \frac{\hat{\rho}_i(\vb{q}_i)}{\vb{q}_i^2} = \frac{e^{2n}}{n!} \qty[\int \!\frac{\dd^3 q}{(2\pi)^3}\frac{1}{\vb{v}'\cdot \vb{q}-\iu 0} \frac{\hat{\rho}(\vb{q})}{\vb{q}^2}]^n~,
\end{equation}
where the equality follows from symmetrizing the eikonal propagators inside the curly braces (introducing the factor of $1/n!$), and using the soft-photon identity \cite{Yennie:1961ad,Peskin:1995ev}, 
\begin{equation}
    \sum_{\rm permutations} \frac{1}{x_{\pi(1)}} \frac{1}{x_{\pi(1)} + x_{\pi(2)} } \ldots \frac{1}{x_{\pi(1)} + x_{\pi(2)}+ \ldots + x_{\pi(n)} } = \frac{1}{x_1} \frac{1}{x_2} \ldots \frac{1}{x_n}~. 
\end{equation}
\end{widetext}

\pagebreak

\vfill
\pagebreak

\bibliographystyle{apsrev4-1}
\bibliography{biblio}

\end{document}